\documentclass[aps,prl,superscriptaddress,twocolumn,amsmath,amssymb]{revtex4}
\usepackage{graphicx}
\begin{document}
\title{Balance between information gain and reversibility in weak measurement}
\author{Yong Wook Cheong}

\affiliation{Gyeonggi Science High School, Suwon, Gyeonggi-Do
440-210, Korea}

\author{Seung-Woo Lee}

\email{swleego@gmail.com} \affiliation{Center for Macroscopic
Quantum Control, Department of Physics and Astronomy, Seoul National
University, Seoul, 151-742, Korea}

\date{\today\\}

\begin{abstract}
We derive a tight bound between the quality of estimating a quantum
state by measurement and the success probability of undoing the
measurement in arbitrary dimensional systems, which completely
describes the tradeoff relation between the information gain and
reversibility. In this formulation, it is clearly shown that the
information extracted from a weak measurement is erased through the
reversing process. Our result broadens the information-theoretic
perspective on quantum measurement as well as provides a standard
tool to characterize weak measurements and reversals.
\end{abstract}
\newcommand{\bra}[1]{\left<#1\right|}
\newcommand{\ket}[1]{\left|#1\right>}
\newcommand{\abs}[1]{\left|#1\right|}
\maketitle

%\section{intro}

Since Heisenberg discussed the $\gamma$-ray microscope
\textit{gedanken} experiment \cite{heisen}, the disturbance induced
by measurement becomes one of the fundamental issues in quantum
mechanics. A heuristic statement, `the more information is obtained
from a quantum system, the more its state is disturbed by
measurement' is widely believed nowadays, and so numerous efforts
have been devoted to prove this in a quantitative manner
\cite{banaszek, banaszek02, andersen,sacchi,dariano}.

However, the general belief in irreversibility of quantum
measurement has been shown to be not always true in the sense that
the input state can be retrieved with a nonzero success probability
by reversing operations on the post-measurement state
\cite{Ueda1992,Koashi1999}. This is because the quantum state is not
fully perturbed by measurement when the interaction between the
system and measurement apparatus is weak. The measurement that
induces a partial collapse of quantum state is called ``weak
measurement'', and its reversing process has been studied
theoretically \cite{Korotkiv2006,Andrew2010,Sun2009} and realized
experimentally \cite{Katz2008,YSKim2009}. It has attracted much
attention due to its potential applications in quantum information
processing \cite{Korotkov2010,Kim2012}.

In information-theoretic point of view, reversibility can be
understood as a degree of preserved information in measurement
process, and thus should be quantitatively related to the extracted
one \cite{nielsen}. In fact, the information that is not extracted
from a measurement is transferred to the remainder of the whole
Hilbert space describing the measurement process (see appendix).
Even after some information is extracted through a measurement, its
state can be retrieved with a probability equal to the degree of
ignorance of the state \cite{Koashi1999}. This concept was also
proved in another context as the `no-hiding theorem' of information
\cite{braunstein}. In this sense, the extracted information should
be more tightly related to the possibility of undoing the
measurement \cite{dariano} rather than the closeness between input
and post-measurement states as used in previous works
\cite{banaszek}. Recently, an entropic trade-off relation was
derived based on the concept of information conservation in
measurement process \cite{Buscemi2008}, and a degree of information
gain was investigated by changing the reversibility in a single
measurement outcome level \cite{Terashima2011}. However, a clear and
direct quantitative relation between information gain and
reversibility in quantum measurement has so far been missing.

In this Letter, we derive a tight bound between the amount of
information gain and reversibility in arbitrary $d$-level systems,
which are quantified by the average estimation fidelity
\cite{banaszek} and the reversal probability \cite{Koashi1999},
respectively. In particular, it shows a sharp trade-off relation
between them with a monotonic equation for qubit (2-level) systems.
To our knowledge, this is the first direct and quantitative link
between information gain and reversibility. Moreover, since both the
estimation fidelity \cite{Genoni2006,andersen,Baek2008} and reversal
probability \cite{Koashi1999,Katz2008,YSKim2009,Sun2009} are
measurable quantities, its demonstration is experimentally feasible.
Our result provides a fundamental insight on the quantum measurement
as well as a useful tool to characterize reversals of weak
measurements potentially used in quantum information processing
\cite{Korotkov2010,Kim2012}.

{\em Quantum measurement --} An ideal measurement can be described
by a set of operators $\{ \hat{A}_r|r=1,\ldots ,N \}$, satisfying
the completeness relation
\begin{equation}
\label{eq:comp} \sum_{r=1}^{N}\hat{A}_r^{\dagger}
\hat{A}_r=\hat{\openone},
\end{equation}
where the index $r$ indicates the obtained classical information. A
measurement performed on a system transforms its input state
$\ket{\psi}$ to
\begin{equation}
\label{eq:redstate}
\ket{\psi_r}=\frac{\hat{A}_r\ket{\psi}}{\sqrt{p(r,\ket{\psi})}},
\end{equation}
which is the {\em post-measurement} state, where
$p(r,\ket{\psi})=\bra{\psi}\hat{A}_r^{\dagger}\hat{A}_r\ket{\psi}$
is the probability that the outcome is $r$.

A measurement operator $\hat{A}_r$ can be written by the
singular-value decomposition:
$\hat{A}_r=\hat{V}_r\hat{D}_r\hat{U}_r$, where $\hat{U}_r$ and
$\hat{V}_r$ are unitary operators, and $\hat{D}_r$ is a diagonal
matrix with non-negative entries. We assume $\hat{V}_r=\openone$
without loss of generality, and $\hat{U}_r$ can be written by
$\hat{U}_r =\sum_{i=0}^{d-1} \ket{v^r_i} \bra{w^r_i}$ with two
orthonormal bases $\{\ket{v^r_i} | i=0,\ldots,d-1 \}$ and
$\{\ket{w^r_i} | i=0,\ldots,d-1\}$ for $d$-level measurements. The
diagonal matrix can be also written by $\hat{D}_r=\sum_{i=0}^{d-1}
\lambda^r_i\ket{v^r_i} \bra{v^r_i}$, with non-negative diagonal
elements $\lambda^r_i$ (i.e. singular values) put in decreasing
order such that $\lambda^r_0\geq \lambda^r_1\geq\ldots \geq
\lambda^r_{d-1}$. Thus, each measurement operator can be represented
by
\begin{equation}
\label{eq:imeas} \hat{A}_r=\sum_{i=0}^{d-1} \lambda^r_i \ket{v^r_i}
\bra{w^r_i},
\end{equation}
and due to the completeness relation in Eq.~(\ref{eq:comp}) their
singular values $\lambda^r_i$ satisfy
\begin{equation}
\label{ea:compsing} \sum_{r=1}^{N}\sum_{i=0}^{d-1}
(\lambda^r_i)^2=d.
\end{equation}

{\em Information gain --} In order to quantify the obtained
information through a measurement, we employ the estimation fidelity
\cite{banaszek}. When the measured outcome is $r$, one can make a
guess on the input state $\ket{\psi}$ and select a state
$\widetilde{\ket{\psi_r}}$. The quality of the guess can be
quantified with the help of overlap between them $|\langle \psi
\widetilde{\ket{\psi_r}}|^2$. Then, the mean estimation fidelity is
obtained by averaging $|\langle \psi \widetilde{\ket{\psi_r}}|^2$
over all possible measurement outcomes $r$ and input states
$\ket{\psi}$:
\begin{equation}
\label{efidelity} G = \int d \psi \sum_{r=1}^{N} p(r,\ket{\psi}) |
\langle \widetilde{\psi_r} | \psi \rangle|^2,
\end{equation}
which gives different values depending on the guess strategy. We
reformulate it by
\begin{equation}
\sum_{r=1}^{N} \int d  \psi \bra{\psi} \otimes \bra{\psi}
(\hat{A}_r^\dagger \hat{A}_r \otimes
\widetilde{\ket{\psi_r}}\widetilde{\bra{\psi_r}})\ket{\psi} \otimes
\ket{\psi},
\end{equation}
and use the Schur's lemma \cite{albeverio} that leads to the
identity,
\begin{eqnarray}
\label{eq:schur} \int_{G} dg
\left[\hat{U}^{\dag}(g)\otimes\hat{U}^{\dag}(g) \right] \hat{O}
\left[\hat{U}(g)\otimes\hat{U}(g) \right] = \alpha_{1}
\hat{\openone}\otimes\hat{\openone} + \alpha_{2} \hat{S},\nonumber \\
\nonumber
\alpha_{1}=\frac{d^2\mathrm{Tr}(\hat{O})-d\mathrm{Tr}(\hat{O}\hat{S})}{d^2(d^2-1)}
,~\alpha_{2}=\frac{d^2\mathrm{Tr}(\hat{O}\hat{S})-d\mathrm{Tr}(\hat{O})}{d^2(d^2-1)},~~
\end{eqnarray}
for any operator $\hat{O}$ acting on the $d\times d$ Hilbert space.
Here $dg$ is Haar invariant measure on the $d$-dimensional unitary
group $G = \mbox{U}(d)$ such that $\int_{G} dg = 1$, $\hat{U}(g)$ is
an irreducible unitary representation of $g \in G$, and $\hat{S}$ is
a swap operator defined as $\hat{S} |i\rangle \otimes \ket{j} =
|j\rangle \otimes \ket{i}$. A simpler form is then obtained as
\begin{equation}
\label{eq:sum} \frac{1}{d(d+1)}(d+\sum_{r=1}^{N}
\mathrm{Tr}[\hat{A}_r^\dagger \hat{A}_r \otimes
\widetilde{\ket{\psi_r}}\widetilde{\bra{\psi_r}}\hat{S}]),
\end{equation}
and by using Eq.~(\ref{eq:imeas}) its second term is rewritten by
\begin{equation}
\sum_{r=1}^N \sum_{i=0}^{d-1}
(\lambda_i^r)^2|\widetilde{\bra{\psi_r}} w_i^r \rangle |^2,
\end{equation}
which gives a maximum value when the estimated state
$\widetilde{\ket{\psi_r}}$ is equivalent to $\ket{w_0^r}$. Then, we
define the measure of {\em information gain} as the maximal value of
the mean estimation fidelity,
\begin{equation}
\label{eq:estifide} G_{max}=\frac{1}{d(d+1)}(d+\sum_{r=1}^N
(\lambda_0^r)^2),
\end{equation}
which is a function of the maximal singular value $\lambda_0^r$ of
the measurement operators. Note that it is scaled in the range $1/d
\leq G_{max} \leq 2/(d+1)$, where the upper bound $2/(d+1)$ is
reachable by a von Neumann measurement and the lower bound $1/d$ is
obtained by a unitary measurement or equivalently by a random guess.
The result in Eq.~(\ref{eq:estifide}) is valid for arbitrary input
states $\hat{\rho}$ as a mixed state degrades the estimation
fidelity by averaging over the input probability so that its maximum
is always obtained in the space of pure states.

{\em Reversibility --} A reversing operator $\hat{R}^{(r)}$ can be
defined for a physically reversible measurement $\hat{A}_r$
\cite{Koashi1999} to recover the input state as
$\hat{R}^{(r)}\ket{\psi_r}\propto \ket{\psi}$. Thus, a subsequent
measurement of reversing operator $\hat{R}^{(r)}$ after the first
measurement $\hat{A}_r$ leads to a successful reversal,
independently on the input state $\ket{\psi}$, as
\begin{equation}
\label{eq:revop} \hat{R}^{(r)}\hat{A}_r\ket{\psi}=\eta_r\ket{\psi},
\end{equation}
where $\eta_r$ is a nonzero complex number.

Since $\hat{R}^{(r)}$ can be regarded as an element of a complete
measurement set, $\openone - \hat{R}^{(r)\dag}\hat{R}^{(r)}$ is
positive semidefinite and equivalently,
\begin{equation}
\label{eq:upper} \sup_{\ket{\phi}}\bra{\phi}\hat{R}^{(r)
\dag}\hat{R}^{(r)}\ket{\phi}\leq1,
\end{equation}
for arbitrary (normalized) quantum state $\ket{\phi}$.
Simultaneously \cite{Koashi1999},
\begin{eqnarray}
\label{eq:lowersup} \nonumber
\sup_{\ket{\phi}}\bra{\phi}\hat{R}^{(r)
\dag}\hat{R}^{(r)}\ket{\phi}&\geq&
\sup_{\ket{\psi_r}}\bra{\psi_r}\hat{R}^{(r)
\dag}\hat{R}^{(r)}\ket{\psi_r}\\
\nonumber
&=&\sup_{\ket{\psi}}\frac{\bra{\psi}\hat{A}^{\dag}_r\hat{R}^{(r)
\dag}\hat{R}^{(r)}\hat{A}_r\ket{\psi}}{p(r,\ket{\psi})}\\
&=&\frac{|\eta_r|^2}{\inf_{\ket{\psi}}p(r,\ket{\psi})},
\end{eqnarray}
so that $|\eta_r|^2\leq\inf_{\ket{\psi}}p(r,\ket{\psi})$ is
satisfied. As the input state can be written with an arbitrary
orthonormal basis $\{\ket{w_i} | i=0,\ldots,d-1\}$ as
$\ket{\psi}=\sum_{i=0}^{d-1}\alpha_i\ket{w_i}$ where
$\sum_{i=0}^{d-1}|\alpha_i|^2=1$ and the singular values are defined
in decreasing order,
\begin{eqnarray}
\label{eq:limitv}
|\eta_r|^2\leq\inf_{\ket{\psi}}p(r,\ket{\psi})=\inf_{\{\alpha_i\}}|\alpha_i
\lambda^r_i|^2=(\lambda^r_{d-1})^2,
\end{eqnarray}
is obtained when $\alpha_{d-1}=1$ and all other $\alpha_i$ are zero.

Therefore, the reversal probability for each measurement outcome $r$
has the upper limit as
\begin{equation}
\label{eq:reverr}
P_{rev}(r)=|\bra{\psi}\hat{R}^{(r)}\ket{\psi}_r|^2=\frac{|\eta_r|^2}{P(r,\ket{\psi})}
\leq \frac{(\lambda^r_{d-1})^2}{P(r,\ket{\psi})}.
\end{equation}
We then define the {\em reversibility} as the maximal mean value of
reversal probability over all the outcomes $r$ \cite{Andrew2010},
\begin{eqnarray}
\label{eq:revers} P_{rev}= \mathrm{max}
\sum_{r=1}^{N}P_{rev}(r)P(r,\ket{\psi})=\sum_{r=1}^{N}(\lambda^r_{d-1})^2,
\end{eqnarray}
which notably does not depend on the input state $\ket{\psi}$ but is
given as a function of the minimal singular value of measurement
operators, $\lambda^r_{d-1}$. Its maximum value $P_{rev}=1$ is
obtained by a unitary measurement, meaning that the input state can
be deterministically retrieved with appropriate reversing unitary
operation, while the minimum value $P_{rev}=0$ is given by a von
Neumann measurement, implying that full extraction of information
frustrates the reversing process.

Assuming arbitrary mixed input states $\hat{\rho}$, we can obtain
the same reversibility with the form in Eq.~(\ref{eq:revers})
\cite{Andrew2010} as
$\inf_{\hat{\rho}}p(r,\hat{\rho})=\inf_{\hat{\rho}}\mathrm{Tr}[\hat{\rho}\hat{A}_r^{\dagger}\hat{A}_r]
=(\lambda^r_{d-1})^2$ and $P_{rev}(r) \leq
(\lambda^r_{d-1})^2/P(r,\hat{\rho})$ so that
$P_{rev}=\sum_{r=1}^{N}(\lambda^r_{d-1})^2$. Therefore, the result
in Eq.~(\ref{eq:revers}) is valid for arbitrary input states.

%The reversibility in Eq.~(\ref{eq:revers}), defined in the
%assumption of pure input states, is also valid for arbitrary mixed
%input states $\hat{\rho}$ as the limit value of reversal probability
%in the space of all possible states coincides with that obtained for
%the pure states only \cite{Andrew2010} {\em i.e.}
%$\inf_{\hat{\rho}}p(r,\hat{\rho})=\inf_{\hat{\rho}}\mathrm{Tr}[\hat{\rho}\hat{A}_r^{\dagger}\hat{A}_r]
%=(\lambda^r_{d-1})^2$ and $P_{rev}(r) \leq
%(\lambda^r_{d-1})^2/P(r,\hat{\rho})$ so that
%$P_{rev}=\sum_{r=1}^{N}(\lambda^r_{d-1})^2$.

It may be considerable to quantify the disturbance of quantum states
by using the reversibility of measurement. For instance, we can
define a measure of disturbance by the quantity $1-P_{rev}$. It
shows that the higher the reversal probability is, the less the
state is disturbed, which satisfies the requirements for measures of
state disturbance listed in Ref.~\cite{dariano}.

{\em Trade-off Relation --} We now derive a trade-off relation
between the information gain and reversibility from the
representation obtained above. An inequality
\begin{equation}
\label{eq:relate}
\sum_{r=1}^{N}\{(\lambda^r_0)^2+(d-1)(\lambda^r_{d-1})^2\} \leq d,
\end{equation}
is derived from the completeness relation in Eq.~(\ref{ea:compsing})
and the non-increasing order of the singular values
($\lambda^r_0\geq \lambda^r_1\geq\ldots \geq \lambda^r_{d-1}$). From
the Eq.~(\ref{eq:estifide}), (\ref{eq:revers}) and
(\ref{eq:relate}), we can finally obtain a bound inequality for
$G_{max}$ and $P_{rev}$ as
\begin{equation}
\label{eq:balance} d(d+1)G_{max}+(d-1)P_{rev} \leq 2d,
\end{equation}
where $1/d \leq G_{max} \leq 2/(d+1)$, which is the main result of
this letter, showing a trade-off relation between information gain
and reversibility.

We can find a measurement that is maximally reversible for a fixed
amount of information gain, which saturates the inequality
Eq.~(\ref{eq:balance}). The necessary and sufficient condition to
reach the equality sign is that each measurement operator has the
form satisfying $\hat{A}^{\dagger}_r\hat{A}_r = a_r\ket{w^r_0}
\bra{w^r_0}+ b_r\hat{\openone}$ for certain nonnegative parameters
$a_r$ and $b_r$. It is thus guaranteed that the inequality in
Eq.~(\ref{eq:balance}) is tight and can not be further improved.
Interestingly, the maximal reversibility in our result does not
necessarily correspond to the minimal disturbance, which is defined
by the closeness of transformed state from the input state $\int d
\psi \sum_{r=1}^{N} |\bra{\psi}\hat{A}_r \ket{\psi}|^2$
\cite{banaszek}, while the converse is true. This implies that our
trade-off relation differs from the one proposed by Banaszek
\cite{banaszek}.

For qubit (2-level) systems, a particulary interesting trade-off
relation is obtained. In this case, the inequality of
Eq.~(\ref{eq:balance}) is reduced to a monotonic equation
\begin{equation}
\label{eq:2tradeoff} 6G_{max}+P_{rev}=4,
\end{equation}
where $1/2 \leq G_{max} \leq 2/3$. We emphasize that $G_{max}$ and
$P_{rev}$ for any ideal measurement should satisfy this equation.
Therefore, we come to a heuristic statement about quantum
measurement `the more information is obtained from quantum system,
the less possible it is to retrieve the input state of the system'.

{\em Erasing Information -- } The trade-off relation
(\ref{eq:balance}) and (\ref{eq:2tradeoff}) implicate the
possibility of erasing information by reversing operation. One may
ask whether it is possible to erase the information already obtained
and possibly recorded somewhere else. The answer is `yes' for any
partial information obtained by weak measurement, while any full
information by von Neumann measurement is not erasable. In order to
describe the erasing process, we will consider two weak
measurements, saying $\{\hat{A}_r\}$ and $\{\hat{B}_\mu\}$,
performed one after the other on an unknown system. Then, the
erasure of information is simply understood as a collection of the
opposite information by $\{\hat{B}_\mu\}$ that makes the information
already obtained by $\{\hat{A}_r\}$ less certain \cite{Andrew2010}.

Let assume that one element of the second measurement set is given
by $\hat{B}_1=\hat{R}^{(r)}$. If the results of two measurements are
given in turn as $r$ and $1$, the total measurement operation
performed on the state is described by
$\hat{B}_1\hat{A}_r=\hat{R}^{(r)}\hat{A}_r$. From
Eq.~(\ref{eq:revop}), it satisfies
$\hat{R}^{(r)}\hat{A}_r\ket{\psi}=\eta_r\ket{\psi}$ independently on
the input state $\ket{\psi}$, meaning that no information is
obtained about the state. Therefore, we conclude that the
information obtained through a measurement is erased by its
reversal.

Since a measurement operator $\hat{A}_r$ is decomposable into
$\hat{A}_r=\hat{D}_r\hat{U}_r$, its optimal reversing operation is
given from Eq.~(\ref{eq:revop}) as
$\hat{R}^{(r)}=\eta_r\hat{U}^{\dag}_r\hat{D}^{-1}_r$ where
$\hat{U}^{\dag}_r=\sum_{i=0}^{d-1} \ket{w^r_i} \bra{v^r_i}$ and
$\hat{D}^{-1}_r=\sum_{i=0}^{d-1} \frac{1}{\lambda^r_i} \ket{w^r_i}
\bra{w^r_i}$, with an assumption that each $\lambda^r_i$ is nonzero.
Then, we can define the {\em erasing operator} for an arbitrary
measurement operator $\hat{A_r}$ as
\begin{equation}
\label{eq:eop}
\hat{E}^{(r)}=\lambda^r_{d-1}\hat{D}^{-1}_r=\sum_{i=0}^{d-1}
\frac{\lambda^r_{d-1}}{\lambda^r_i} \ket{w^r_i} \bra{w^r_i}.
\end{equation}
It transforms the post-measurement state $\ket{\psi_r}$ to
\begin{equation}
\label{eq:unirev} \hat{E}^{(r)}\ket{\psi_r}=\sqrt{P_{er}}
\hat{U}_r\ket{\psi},
\end{equation}
where $P_{er}=\bra{\psi_r}\hat{E}^{(r)}\hat{E}^{(r)}\ket{\psi_r}
=(\lambda^r_{d-1})^2/P(r,\ket{\psi})$, from which the input state
$\ket{\psi}$ can be retrieved deterministically by unitary operation
\cite{nielsen}, meaning that at this stage the information obtained
by $\{\hat{A_r}\}$ is erased.

{\em Examples -- } (i) Assume the case when a von Neumann
measurement with two operators $\hat{A}_1=\ket{0}\bra{0}$ and
$\hat{A}_2=\ket{1}\bra{1}$ is performed on an arbitrary qubit. Then,
the degree of information gain has the maximal value $G_{max}=2/3$
with a zero reversibility ($P_{rev}=0$) irrespectively on the input
state. It shows that the von Neumann measurement can not be reversed
in any case (the information can not be erased).

(ii) Consider a weak measurement described by two operators
$\hat{A}_1=\sqrt{\eta}\ket{1}\bra{1}$ and
$\hat{A}_2=\ket{0}\bra{0}+\sqrt{1-\eta}\ket{1}\bra{1}$ where $\eta$
is defined as the probability of detecting $\ket{1}$ state (as
implemented in Ref.~\cite{YSKim2009}). If the measurement outcome is
$r=1$ the state collapses on the state $\ket{1}$, while when $r=2$
the input state collapses partially and can be retrieved. The degree
of information gain is $G_{max}=(3+\eta)/6$ and the reversibility is
$P_{rev}=1-\eta$, satisfying the trade-off relation
(\ref{eq:2tradeoff}).

The information obtained by this measurement can be erased by
properly chosen another measurement. From Eq.~(\ref{eq:eop}), the
erasing operator for $\hat{A}_2$ (where
$\hat{D}^{-1}=\ket{0}\bra{0}+(1/\sqrt{1-\eta})\ket{1}\bra{1}$ and
$\lambda_1^{2}=\sqrt{1-\eta}$) is given as
\begin{eqnarray}
\hat{E}^{(2)}&=&\lambda_1^{2}\hat{D}^{-1}=\sqrt{1-\eta}\ket{0}\bra{0}+\ket{1}\bra{1}.
\end{eqnarray}
A measurement $\{\hat{B}_{\mu}|\mu=1,2\}$ can be then defined with
two operators $\hat{B}_1=\hat{E}^{(2)}$ and
$\hat{B}_2=\sqrt{\eta}\ket{0}\bra{0}$, satisfying the completeness
relation $\hat{B}^2_1+\hat{B}^2_2=\openone$. Thus, the information
extracted from the result $r=2$ of the first measurement
$\{\hat{A}_r\}$ is erased probabilistically by the subsequent
measurement $\{\hat{B}_\mu\}$ when its outcome is $\mu=1$, since the
result of second measurement makes the information obtained from the
first measurement uncertain. Our formalism is generally applicable
to any examples of weak measurements and reversals in
Ref.~\cite{Korotkiv2006,Andrew2010,Sun2009,Katz2008,YSKim2009,Korotkov2010,Kim2012}.

{\em Remarks --} Our result provides a useful framework to
generalize the quantum teleportation \cite{bennett93}. Suppose that
Alice performs a joint measurement (assumed here as a projection $\{
\ket{w^r}_{ab}\bra{w^r}\}$ for simplicity) on an unknown input
$\ket{\psi}_a$ and one party of an entangled channel
$\ket{\Psi}_{bc}$. Here $a$, $b$ and $c$ denote the input, Alice's
and Bob's modes, respectively. The teleportation can be then
described as a reversible measurement with operators $\{
{}_{ab}\langle w^r |\Psi \rangle_{bc} \}$ performed on
$\ket{\psi}_a$ so that, based on our formalism, the extracted
information of $\ket{\psi}_a$ during the teleportation and its
reversibility are certainly in the trade-off relation. As the
reversibility here indicates the success probability of
teleportation, the result is rephrased as `the less information
about input state is disclosed during the teleportation, the higher
the teleportation probability'. For example, the standard
teleportation \cite{bennett93} is deterministic as Alice cannot
obtain any information of $\ket{\psi}_a$ by the Bell measurement
with a maximally entangled channel. Within this framework, various
tasks of quantum transmission ({\em e.g.} from the teleportations
using non-maximally entangled or non-orthogonal measurements with
arbitrary entangled channels to the communications in quantum
networks \cite{qnetwork}) can be characterized. The detailed
analysis of generalized teleportation will be presented elsewhere.

Obviously our result manifests the quantum no-cloning theorem in
information-theoretic perspective \cite{wootters82}, as a perfect
copy of a quantum state would violate the bound in
Eq.~(\ref{eq:balance}), which is a crucial ingredient of quantum
cryptography \cite{gisin2002}. Another implication of our result is
that the success rate of quantum error correction should be bounded
by the amount of information loss in qubit \cite{Koashi1999}, which
may lead to further applications in quantum computation.

In summary, we derive a trade-off relation between the degree of
information gain and reversibility in arbitrary-dimensional quantum
measurement. It quantitatively shows that {\em `the more information
is obtained from quantum measurement, the less possible it is to
undo the measurement'}. Simultaneously, it is clearly shown that
undoing a quantum measurement erases the same amount of information
obtained by the measurement. Our result, as providing an
information-theoretic insight on quantum measurement, is expected to
widen the potential applications of weak measurements and reversals
in quantum information processing.

\begin{acknowledgments}
We thank J. Lee and H.-W. Lee for discussions. Y.W.C. acknowledge
supports from the National Research Foundation of Korea Grant funded
by the Korea government (Grant No. NRF-2009-351-C00028). S.W.L.
acknowledge supports from the National Research Foundation of Korea
(NRF) funded by the Korea government (Grant No. 3348-20100018), and
the T. J. Park Foundation.
\end{acknowledgments}

{\em Appendix --} %We shall prove the fact that {\em a complete
%ignorance about the measured system manifests the preservation of
%total information in the measurement process}. For simplicity, we
%consider here a qubit system (the result can be easily extended to
%$d$-level system).
Suppose that an arbitrary input state
$\ket{\psi}= \alpha \ket{0}+\beta \ket{1}$ and ancillary $n$-qubit
states $\left| 0 \right\rangle ^{ \otimes n}$ are prepared for the
measurement. A general measurement can be described as the
combination of an unitary operation $U$ acting on the total
$(n+1)$-qubits and a projection measurement acting on the selected
$m$-qubits out of the $(n+1)$-qubits. The probability that
$m$-qubits are projected on $\hat{P}_{\bar i}  = \left| {\bar i}
\right\rangle \left\langle {\bar i} \right| = \left| {i_1
,\ldots,i_m } \right\rangle \left\langle {i_1 ,\ldots,i_m } \right|
(i_1,\ldots,i_m \in \{0,1\}) $ is given by
\begin{equation}
\label{eq:appen1} p_{\bar i}  = \mathrm{Tr}\left( {\hat{P}_{\bar
i}\hat{U}\left| \psi \right\rangle \left\langle \psi  \right|
\otimes \left| 0 \right\rangle \left\langle 0 \right|^{ \otimes n}
\hat{U}^\dag \hat{P}_{\bar i} }
  \right).
\end{equation}
If the probability $ p_{\bar i}$ of each measurement outcome $\bar
i$ is independent on the input state $\ket{\psi}$, then no
information about $\ket{\psi}$ is obtained through the measurement.
In this case, the input state can be retrieved deterministically as
shown below.

Let us define  $| {\psi _j }\rangle=\hat{U}\left| j \right\rangle
\otimes \left| 0 \right\rangle ^{ \otimes n} $ ($j \in \{ 0,1\}$).
Since the probability $p_{\bar i}$ in Eq.~(\ref{eq:appen1}) is
invariant for any input state $\ket{\psi}$, we obtain an orthogonal
condition
\begin{equation}
\label{eq:appen2} {\langle {\psi _0 } | {\bar i} \rangle~  \cdot
\langle {\bar i} | {\psi _1 } \rangle }  = 0,
\end{equation}
where $\langle {\psi _0 } | {\bar i} \rangle$ is a $(n-m+1)$-qubit
bra, and the symbol $\cdot$ denotes an inner product. By normalizing
${\langle {\bar i}} | {\psi _0 } \rangle $ and ${\langle {\bar i}} |
{\psi _1 } \rangle $, we obtain two orthonormal vectors, saying
$|\varphi _{0 \bar{i}}\rangle$ and $|\varphi _{1 \bar{i}}\rangle$.
Then $| {\psi _j } \rangle$  can be represented by
\begin{eqnarray}
\left| {\psi _j } \right\rangle  = \sum\limits_{\bar i} {\sqrt
{p_{\bar{i}} } \left| {\bar i} \right\rangle  \otimes \left|
{\varphi _{j\bar{i}} } \right\rangle } ,
\end{eqnarray}
where $| {\bar i}\rangle$ and $|\varphi _{ji}\rangle$ are projected
$m$-qubit state and  corresponding $(n-m+1)$-qubit state,
respectively. As $\hat{U}$ is a linear operator, the evolution of
total $n+1$-qubits under $\hat{U}$ is given by
\begin{equation}
\hat{U}\left| \psi  \right\rangle \otimes \left| 0 \right\rangle ^{
\otimes n}  = \sum\limits_i {\sqrt {p_i } \left| {\bar i}
\right\rangle \otimes \left( {\alpha \left| {\varphi _{0\bar{i}} }
\right\rangle  + \beta \left| {\varphi _{1\bar{i}} } \right\rangle }
\right)}.
\end{equation}
If the outcome on $m$-qubit projection is $ | {\bar i} \rangle$,
then remaining $(n-m+1)$-qubits are reduced to $ \alpha | \varphi
_{0\bar{i}} \rangle + \beta | \varphi _{1 \bar{i}} \rangle $. Since
$ |\varphi _{0 \bar{i}}\rangle$ and $|\varphi _{1\bar{i}}\rangle $
are orthonormal vectors determined by $\hat{U}$, we can retrieve the
input state by acting proper unitary operation on the remaining
state. The reversal is possible for any measurement outcome $
\bar{i}$ whenever $p_{\bar{i}}$ is independent of the input state.
We thus conclude that if no information is extracted through the
measurement, the whole information is preserved in the remaining
part of the Hilbert space describing the measurement and the
original state can be retrieved deterministically.

\end{document}